# FastPacket: Towards Pre-trained Packets Embedding based on FastText for next-generation NIDS

Khloud Al Jallad*

*Correspondence: k-aljallad@aiu.edu.sy

Faculty of Information Technology, Arab International University, Daraa, Syria

______________________________________________________________

## ABSTRACT

New Attacks are increasingly used by attackers everyday but many of them are not detected by Intrusion Detection Systems as most IDS ignore raw packet information and only care about some basic statistical information extracted from PCAP files. Using networking programs to extract fixed statistical features from packets is good, but may not enough to detect nowadays challenges. We think that it is time to utilize big data and deep learning for automatic dynamic feature extraction from packets. It is time to get inspired by deep learning pre-trained models in computer vision and natural language processing, so security deep learning solutions will have its pre-trained models on big datasets to be used in future researches. In this paper, we proposed a new approach for embedding packets based on character-level embeddings, inspired by FastText success on text data. We called this approach FastPacket. Results are measured on subsets of CIC-IDS-2017 dataset, but we expect promising results on big data pre-trained models. We suggest building pre-trained FastPacket on MAWI big dataset and make it available to community, similar to FastText. To be able to outperform currently used NIDS, to start a new era of packet-level NIDS that can better detect complex attacks.

______________________________________________________________

# 1. Introduction

A pre-trained model is a saved network that was previously trained on a big dataset, we can either use the pre-trained model as is or use transfer learning to customize this model to a given task. The intuition behind transfer learning is that if a model is trained on a large and general enough dataset, this model will effectively serve as a generic model. We can take advantage of these learned feature maps without having to start from scratch by training a large model on a large dataset.

Although its great success in natural language processing [1] and image processing [2], pre-trained models are not yet used in security. In this paper, we suggest to start a new era of security solutions based on pre-trained models.

Till now, IDS heavily rely on the discrete handcrafted features, while deep learning automatic features based on n-grams of raw pcaps may be better solutions to detect complex attacks.

## 1.1. Network IDS Hierarchy

NIDS types based on the data source that it is monitoring are shown in figure 1.

- **Log-based NIDS**: that analyzes logs written by security devices when packets flow.
- **Raw Data-based NIDS** that analyzes the data sent itself, it has two types
    - **Traffic-based**: also called Deep Packet Inspection (DPI), it contains the whole packets' data, headers and bodies.
    - **Flow-based**: it contains headers of packets only.

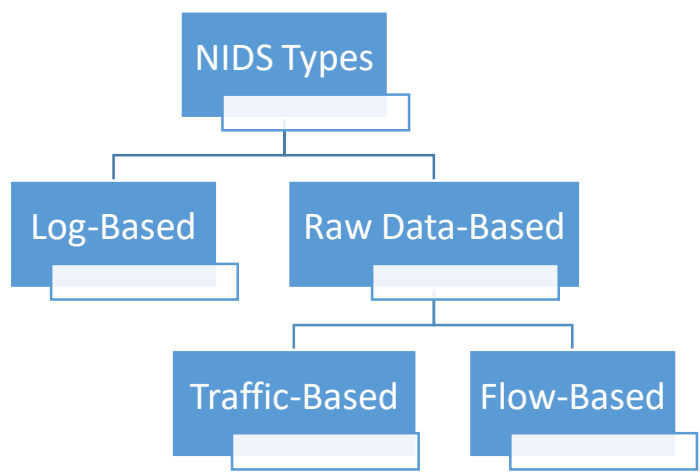

**Figure 1 NIDS Types [3]**

As for Traffic-based (packet-level NIDS), it needs high cost for huge computational resources and not done easily on high-speeds networks, especially when it comes to big data networks (more than 1 TB in second), it is considered time-consuming. The problem of resource demand causes many researchers to use flow-based NIDS.

As for Flow-based (flow-level NIDS), it is very lightweight and useful in lots cases but has high false positive. Storage issues that appeared in the packet-based approach are almost disappeared, but some types of attacks that are injected in bodies cannot be detected by analyzing headers only. [4]

Flow-Level NIDS needs less performance cost than Packets-Level NIDS, However, lots of critical information are dropped when body of packet is ignored, especially information related to some critical attacks, such as, SQL injection attacks, phishing attacks and Trojans that seems normal when only analyzing header information.

In general, Flow-level NIDS uses anomaly-based detection methods and packet-level NIDS uses signature-based detection methods. Each type has its advantages and disadvantages. Therefore, we need to tradeoff high cost or high false positive. Some researchers believe that a combination of both is the best solution. [4] For that reason, we have done this paper.

In this research we propose to combine them in a pre-trained model to overcome the cost and time-consuming process of training and get the advantage of better accuracy.

This paper is organized as follows, we will talk about related works in "Related Work Section". The proposed method is explained in detail in "Method Section". "Data Section" contains detailed information about used dataset. We will discuss results in "Results and Discussion Section". We will talk about conclusion and future vision in "Conclusion and Future Work Section".

## 2. Related Work

Most recent previous works on NIDS used SVM and LSTMs on hand-crafted features for anomaly detection NIDS. [5]

Few recent articles trying to embed packets inspired by famous NLP Word-Embedding.

LogBERT [6] for analyzing anomaly logs based on BERT for text. They have extracted some regular expressions from logs then embed results using BERT embeddings.

Packet2Vec [7] utilized w2vec for packets embedding on DARPA2009 dataset and their results outperformed state-of-the-art results on same dataset.

Both papers are great and inspired us to do this article. But, although both used embedding, no encoding was done before embedding to reduce alphabet size thus processing cost, Moreover Applying NLP famous embedding is a great idea but we think that it is very expensive to always start from scratch when it comes to the cost of processing and storage. We argue that it is essential to have security special pre-trained anomaly models, not only train NLP embedding on a dataset used for research.

## 3. Data

The dataset used for this experiment is CIC-IDS-2017 dataset [8] that contains benign and some common attacks, which resembles the true real-world data (PCAPs). The data capturing period started at 9 a.m., Monday, July 3, 2017 and ended at 5 p.m. on Friday July 7, 2017, for a total of 5 days. Monday is the normal day and only includes the benign traffic. The implemented attacks include Brute Force FTP, Brute Force SSH, DoS, Heartbleed, Web Attack, Infiltration, Botnet and DDoS. They have been executed both morning and afternoon on Tuesday, Wednesday, Thursday and Friday. Some basic information of data is shown in table1

Table 1 CIC-IDS-2017 Dataset Basic Information

| Day | Description | Size (GB) |
| --- | --- | --- |
| Monday | Normal Activity | 11.0G |
| Tuesday | Attacks + Normal Activity | 11G |
| Wednesday | Attacks + Normal Activity | 13G |
| Thursday | Attacks + Normal Activity | 7.8G |
| Friday | Attacks+ Normal Activity | 8.3G |

However, the dataset we propose to be used in pre-trained model is MAWI dataset [9]MAWI stands for Measurement and Analysis on the WIDE Internet [10] [11] [12]. It is the biggest online available and most real-life dataset that is publically available for free. It contains real-life traffic data of Japan-US cable. It is collected and preprocessed by a sponsor of the Japanese ministry of communication

MAWI is labeled by MAWILAB project [13] Which is a project done on top of MAWI archive MAWILAB contains labels of MAWI data, and it is automatically updated daily. Labeling data is done by four classifiers. Classifiers are Principal component analysis (PCA), Gamma Distribution, Hough Transform, Kullback–Leibler (KL). Labels are tagged according to class of majority classifiers detection. That help reducing false positive rate. Labels of MAWILAB are done according to taxonomy of anomalies in network traffic [14]

**4. Proposed Method**

Instead of creating hand-crafted features for each packet, we proposed to encode raw packet data then embed it using FastText vectorization for each packet, then perform classification.

Specifically, our approach has the following steps: We proposed encoding packets' raw content in hex decimal encoding to have limited character alphabets for character-level embeddings. Then we do FastText supervised learning on hex-encoding to build a model that can take n-grams of raw content into consideration as many of attacks are injected in packet payload, such as SQL injections, Phishing attacks and Trojans. After applying Fasttext embedding, we have applied several traditional machine learning algorithms on subsets of used dataset. We chose traditional machine learning approaches as our experiments are done on small datasets only because of hardware limitations. But we think that pre-trained models will be better used with convolutional or sequential deep models

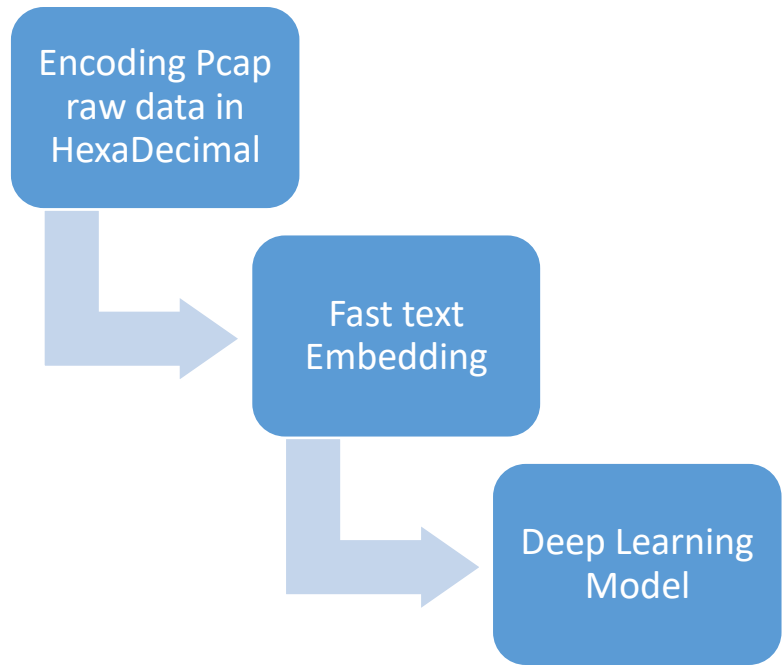

**Figure 2 Proposed Solution**

## 5. Proposed Framework and Libraries

### 5.1. FastText

we build our code based on FastText [15] [16] FastText is an open-source, free, lightweight library that allows users to learn text representations and text classifiers. We have used supervised learning approach; thus an attack pattern can be learned in embeddings.

### 5.2. Scapy

we have used scapy [17] for reading and manipulating packets' raw data. Scapy is a powerful interactive packet manipulation program written in python language.

## 6. Results and Discussion

We chose some traditional machine learning algorithms such as random forest and SVM as it is the most widely used classifier in security researches. We think that building pre-trained model will enable us to get the advantage of higher accuracy by using packet-level NIDS, with avoiding the packet-level NIDS disadvantage of high time and cost needed to train model on each dataset. But our results are not accurate as we compare random subsets of dataset, thus we will not share numbers or charts. But, based on basic research and analysis done on data, we expect promising results on big datasets, therefore we share our insights so other researchers who have better hardware infrastructure can create a pre-trained model dataset and make available to community to be used in future researches.

Raw payload is very important to take into consideration and better than hand-crafted features because many attacks are injected in packet payload, such as SQL injections, Phishing attacks and Trojans.

Encoding raw packets in Hex Encoding can be useful to limit character alphabet size instead of ASCI or UNICODE basic encoding of packets.

We chose traditional machine learning approaches as our experiments are done on small datasets only because of hardware limitations. But we think that big pre-trained models will be better used with convolutional or sequential deep classifiers.

Although the number of hacking attacks is growing exponentially every few months, a general pattern can be automatically extracted by pre-trained anomaly models on big datasets. Same as we see in computer vision and natural language processing that have a great accuracy detection rates despite the wide diversity of texts and images that are growing also exponentially thank to big data pre-trained models.

We recommend using MAWI dataset [9] for pre-trained model as it is the biggest and most organized available online PCAP dataset for anomaly-IDS tasks.

## 7. Conclusion & Future Work

Similar to the huge impact that anomaly-based IDS outperform signature-based IDS in detecting new threats, Automatic deep anomaly pre-trained models are promising to outperform deep hand-crafted anomaly-IDS, because of its ability to detect new threats injected in raw packet contents. We proposed building pre-trained FastPacket models on big datasets, so we can get the advantage of higher accuracy by using packet-level NIDS, with avoiding the packet-level NIDS disadvantage of high time and cost needed to train model on each dataset.

We discuss our claim the reason behind our hypothesis but we were not able to do a complete experiment on a big dataset and create a pre-trained model because of hardware limitations as this type of experiments needs big companies' hardware capabilities to be implemented. The experiments we did on random small subsets of dataset were promising but not enough to prove our hypothesis. For that reason, we share our experiment and our thoughts as we wish that complete experiment can be done in future by interested researchers who have better hardware infrastructure than ours. We wish that a university or a big data company create a pre-trained anomaly model on PCAP files of MAWI big dataset and make it available to community for future research to start a new era of security intelligent solutions.

**Abbreviations**

IDS: Intrusion Detection System,

**Authors' contributions**

KHJ took on the main role so she performed the literature review, conducted the experiments, analyzed results and wrote the manuscript

**Competing interests**

The authors declare that they have no competing interests.

**Acknowledgement**

Not Applicable.

**Availability of data and materials**

All datasets in this survey are available online, you can find links in references.

**Consent for publication**

The authors consent for publication.

**Ethics approval and consent to participate**

The authors Ethics approval and consent to participate.

**Funding**

The authors declare that they have no funding.